\documentclass[twocolumn]{aastex631}

\usepackage{graphicx}	% Including figure files
\usepackage{amsmath}	% Advanced maths commands
\usepackage{hyperref}
\usepackage{cprotect}

\newcommand{\jr}{\color{black}}

\begin{document}

\title{Conclusive evidence for a population of water-worlds around M-dwarfs remains elusive}

\author[0000-0001-7615-6798]{James G. Rogers}
\affiliation{Department of Earth, Planetary, and Space Sciences, The University of California, Los Angeles, 595 Charles E. Young Drive East, Los Angeles, CA 90095, USA}

\author[0000-0002-0298-8089]{Hilke E. Schlichting}
\affiliation{Department of Earth, Planetary, and Space Sciences, The University of California, Los Angeles, 595 Charles E. Young Drive East, Los Angeles, CA 90095, USA}

\author[0000-0002-4856-7837]{James E. Owen}
\affiliation{Astrophysics Group, Department of Physics, Imperial College London, Prince Consort Rd, London, SW7 2AZ, UK}


\begin{abstract}
The population of small, close-in exoplanets is bifurcated into super-Earths and sub-Neptunes. We calculate physically motivated mass-radius relations for sub-Neptunes, with rocky cores and H/He dominated atmospheres, accounting for their thermal evolution, irradiation and mass-loss. For planets $\lesssim 10~$M$_\oplus$, we find that sub-Neptunes retain atmospheric mass fractions that scale with planet mass and show that the resulting mass-radius relations are degenerate with results for `water-worlds' consisting of a 1:1 silicate-to-ice composition ratio. We further demonstrate that our derived mass-radius relation is in excellent agreement with the observed exoplanet population orbiting M-dwarfs and that planet mass and radii alone are insufficient to determine the composition of some sub-Neptunes.
Finally, we highlight that current exoplanet demographics show an increase in the ratio of super-Earths to sub-Neptunes with both stellar mass (and therefore luminosity) and age, which are both indicative of thermally driven atmospheric escape processes. Therefore, such processes should not be ignored when making compositional inferences in the mass-radius diagram.
\end{abstract}

\keywords{planets and satellites: atmospheres -
planets and satellites: physical evolution - planet star interactions}


\section{Introduction} \label{sec:intro}

The observed population of small, close-in exoplanets with radii $\lesssim 4R_\oplus$ and orbital periods $\lesssim 100$ days \citep[e.g.][]{BoruckiKeplerII,Howard2012,Fressin2013,Silburt2015,Mulders2018,Zink2019,Petigura2022} provide an intriguing problem in terms of their formation pathway. With no analogue in our solar system, such planets have been observed to bifurcate into two separate sub-populations, centred at $\sim 1.3 R_\oplus$ (referred to as `super-Earths') and $\sim 2.4 R_\oplus$ (referred to as `sub-Neptunes'). In between, there exists an absence of planets, labelled as the `radius gap', which decreases in size with increasing orbital period \citep[e.g.][]{Fulton2017,VanEylen2018,Berger2020,Petigura2022}. 

Two categories of evolutionary models have emerged to explain this phenomenon. The first relies on atmospheric evolution, as it is known that many sub-Neptunes require a significant H/He dominated atmosphere to explain their observed mass and radius \citep[e.g.][]{Weiss2014,JontofHutter2016,Benneke2019}. Under this class of models, super-Earths are expected to have lost their primordial atmosphere and are thus observed at their core\footnote{The term `core', as used for the remainder of this letter, refers to the solid/liquid bulk interior of a planet, as opposed to the geological nomenclature of the iron core. In general, the mass of a sub-Neptune is approximately given by its core mass, as the atmospheric mass makes up less, often much less, than 10\% of the planet's total mass.} radius. Sub-Neptunes, on the other hand, have maintained their atmosphere, bloating their size to the observed peak at $\sim 2.4 R_\oplus$. Typically, atmospheric mass-loss is thought to cause this bifurcation, with smaller mass, highly irradiated planets losing their hydrogen atmospheres to become super-Earths, whilst larger mass, colder planets remain as sub-Neptunes. Two successful mass-loss models are XUV photoevaporation, which relies on high-energy stellar flux \citep[e.g.][]{Owen2013,LopezFortney2013} and core-powered mass-loss, which calls upon remnant thermal energy from formation and bolometric stellar luminosity \citep[e.g.][]{Ginzburg2018,Gupta2019}. Other models may also explain the radius gap via atmospheric escape due to giant impacts \citep[e.g.][]{Inamdar2016,Wyatt2020} or through gaseous accretion of primordial atmospheres \citep[e.g.][]{LeeConnors2021,Lee2022}. 

An alternative model to the retention of H/He atmospheres is the `water-world' hypothesis, in which the radius gap arises due to a difference in planet composition; super-Earths consisting of a silicate-iron mixture, sub-Neptunes consisting of an ice-silicate mixture. Since a planetary core of a given mass increases in size for lower bulk densities, the water-world/sub-Neptune population exists at a larger size and hence separated from the rocky/iron-rich super-Earths \citep[e.g.][]{Mordasini2009,Raymond2018,Zeng2019,Turbet2020,Mousis2020}.  %In light of the differences between the water-world and the atmospheric-mass loss models, it follows that unlocking the origins of the radius gap is of crucial importance for understanding the underlying formation pathways of small, close-in exoplanets. 

{\jr{These two end-member evolutionary models are not, however, mutually exclusive. Planets are likely to harbour H/He dominated atmospheres as well as water content \citep[i.e.][]{Venturini2016,Lambrechts2019,Mordasini2020,Venturini2020}, which may interact with each other as well as the silicate mantle and iron core of the planet \citep[e.g.][]{Dorn2021,Vazan2022,Schlichting2022}. Whilst the specifics of these interactions remain uncertain, what remains clear is that the key to determining the nature of such planets is to ensure that models are consistent with demographic observations i.e. can reproduce a clean radius gap with the correct orbital period and stellar mass dependencies. This naturally requires a distinct difference in bulk densities between super-Earths and sub-Neptunes.}}

{\jr{In this letter we focus on a specific class of hypothesised water-world, namely that of 1:1 silicate-to-ice ratios in the absence of H/He dominated atmospheres. Such compositions arise as a result of condensation models of solar composition gas at fixed pressure\footnote{Although we note that it remains unclear as to whether a 1:1 composition ratio would appear for condensation models that consider a realistic evolving protoplanetary disc.} \citep[e.g.][]{Zeng2019}. Under such models, volatiles condense beyond their respective ice-line and are speculated to form the constituents of growing planets, typically via pebble accretion. Once planets are sufficiently massive, they migrate inwards towards the typical orbital periods that are observed today \citep[e.g.][]{Lodders2003,Bitsch2015,Venturini2020,Bruger2020}.

The nature of sub-Neptunes in this latter scenario is thus in direct disagreement with alternative models that include a single formation pathway for super-Earths and sub-Neptunes in conjunction with an atmospheric escape model. For example, inferences from photoevaporation and core-powered mass-loss models require that both super-Earths and sub-Neptunes have core bulk densities roughly consistent with that of Earth ($\sim 5.5$ g cm$^{-3}$ for a $1M_\oplus$ core), suggesting that they formed through the same pathway and divided as a result of atmospheric escape \citep[e.g.][]{Wu2019,Gupta2019,Rogers2021,Rogers2023}.}} 

In a recent study, \citet{Luque2022} asserted that there is evidence for a population of the aforementioned water-worlds among planets orbiting M-dwarfs, by comparing observed planet masses and radii with various planet-composition models in the mass-radius diagram. They find that many planets in their sample are consistent with a 1:1 silicate-to-ice composition ratio, as well as population synthesis modelling from \citet{Burn2021}. They also use the mass-radius relations for rocky bodies hosting H/He dominated atmospheres from \citet{Zeng2019} to claim that the planet sample was inconsistent with rocky cores hosting H/He atmospheres. Unfortunately, these adopted mass-radius models for H/He dominated atmospheres were not appropriate for planets at fixed age i.e. analogous to stellar isochrones. In order to do this analysis, one must consider the thermal evolution and atmospheric mass-loss with the associated changes in entropy of H/He atmospheres over time.

{\jr{In this letter, we calculate physically motivated, self-consistent, population-level mass-radius relations for rocky planets hosting H/He atmospheres, which crucially take into account atmospheric evolution and irradiation from the host star. We compare these relations to the data compiled within \citet{Luque2022} in Section \ref{sec:Results}, with discussion and conclusions found in Section \ref{sec:Conclusions}.}}

\section{Method}\label{sec:Method}

It is commonplace for mass-radius diagrams to be used as a visual guide to the population of observed exoplanets \citep[e.g][]{Wu_Lithwick2013,Weiss2014,HaddenLithwick2014,Rogers2015,Dressing2015,Wolfgang2016,Chen2017,VanEylen2021}. To interpret the observations, theoretical mass-radius relations are used to plot a planet's size as a function of mass for a given composition. For solid bodies consisting of iron, silicate and ice mass fractions, the models of \citet{Fortney2007, Zeng2019} are commonly used, in which the planet radius $R_\text{p}$ scales approximately as $R_\text{p}/R_\oplus \propto (M_\text{p} / M_\oplus)^{1/4}$, where $M_\text{p}$ is the planet mass \citep{Valencia2006}. Specifically the models of \citet{Zeng2019}, which were adopted in \citet{Luque2022}, give the following mass-radius relation for water-worlds consisting of a 1:1 silicate-to-ice ratio:
\begin{equation} \label{eq:MR_WW}
    \frac{R_\text{p}}{R_\oplus} \approx 1.24 \; \bigg ( \frac{M_\text{p}}{M_\oplus} \bigg)^{0.27}.
\end{equation}
{\jr{We highlight again that, for the remainder of this letter, we define water worlds as planets with a 1:1 silicate-to-ice ratio, as in \citet{Luque2022}, and thus follow the mass-radius relation given in Eq. \ref{eq:MR_WW}.}} 

Our task is to determine the mass-radius relation for rocky/iron-rich cores with a H/He dominated atmosphere. Crucially, we aim to calculate physically-motivated, self-consistent mass-radius relations, which incorporate the physics of atmospheric evolution, including mass-loss and cooling, which strongly modifies the mass-radius relation from the canonical H/He results of \citet{Zeng2019}. We highlight that the H/He mass-radius models from \citet{Zeng2019} assume constant specific entropy in a purely adiabatic atmosphere. The assumption of constant specific entropy (which sets the adiabat) for planets of varying mass is not accurate for planets that have undergone thermodynamic processes such as cooling and mass-loss, which naturally reduce the specific entropy of a planet and depend on many variables such as planet mass and equilibrium temperature. Moreover, in the \citet{Zeng2019} mass-radius relations, each model's specific entropy is parameterised with a temperature defined at a fixed pressure of 100 bar. We note that this temperature is often mistaken for the equilibrium temperature of a given planet. The temperature and density of a purely adiabatic atmosphere will drop far below the equilibrium temperature within a few scale heights of the planet's surface. In reality, an outer radiative layer will form as a planet comes into radiative equilibrium with the host star \citep[e.g.][]{Guillot2010,Lee2014,Piso2014,Ginzburg2016}. We point the reader to the mass-radius relations of \citet{Lopez2014, ChenRogers2016}, which account for thermal evolution, including radiative-convective models that provide planet size at a constant age for a given mass and H/He mass fraction.

\subsection{Constructing mass-radius relations}

\begin{figure*}
    \centering
        \includegraphics[width=1.9\columnwidth]{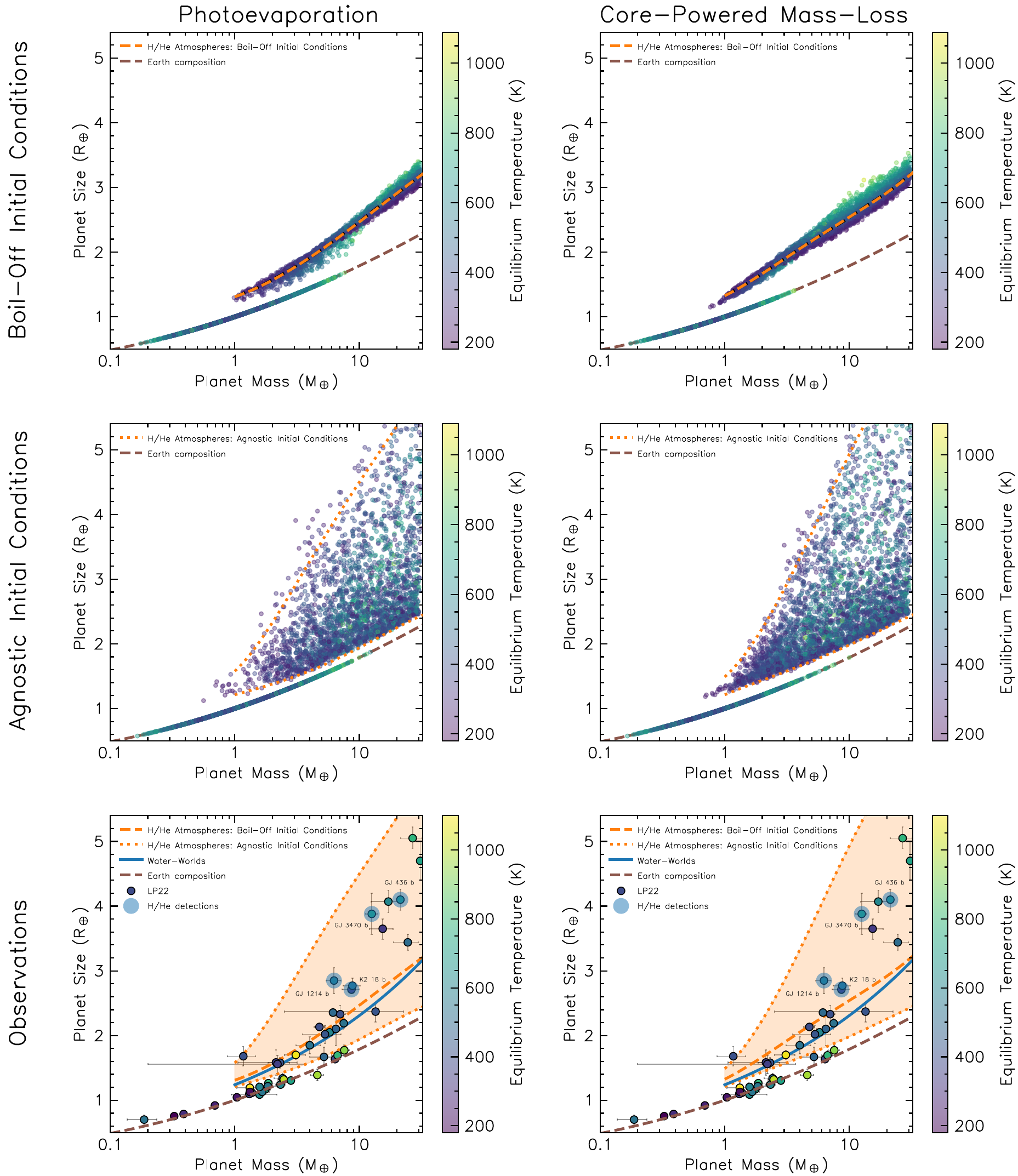}
        \cprotect\caption{Synthetic mass-radius distributions are shown for populations of planets evolved with photoevaporation and core-powered mass-loss in left and right-hand panels, respectively, coloured by their equilibrium temperatures. Super-Earths are stripped of their H/He dominated atmospheres and fall onto a relation consistent with an Earth-like composition (brown-dashed), whilst sub-Neptunes retain a significant atmosphere. In the top panels, we assume an initial distribution of atmospheric masses appropriate for a boil-off scenario (Eq. \ref{eq:Xinit_boiloff}), in which planets lose a significant amount of H/He mass during disc dispersal. We characterise the resulting narrow mass-radius distribution with a median line (orange dashed, Eq. \ref{eq:MR_NEW}). In the middle panels, we adopt agnostic initial conditions (Eq. \ref{eq:Xinit_agnostic}) and parameterise this mass-radius relation with $2\sigma$ limits (orange dotted lines). In the bottom panels, we compare our theoretical mass-radius distributions (orange dashed/dotted lines, Eq. \ref{eq:MR_NEW}) with the observed sample of M-dwarf orbiting exoplanets from \citet{Luque2022}, together with the mass-radius relation for water-worlds (blue solid line, Eq. \ref{eq:MR_WW}). We find that boil-off initial conditions provide mass-radius relations that are completely degenerate with that of water-worlds. Furthermore, even when adopting agnostic initial conditions, the observations are accurately reproduced since the mass-radius distribution is naturally explained due to mass-loss and cooling/contraction of H/He dominated atmospheres. We highlight planets with confirmed escaping H/He detections with blue-shaded regions (namely; K2 18 b, GJ 3470 b, GJ 436 b and, tentatively, GJ 1214
        b).} \label{fig:MR_combined} 
\end{figure*} 

Atmospheric mass-loss sculpts the exoplanet population such that planets with larger core masses and therefore deeper gravitational potential wells retain larger atmospheric masses. This is also true for planets at cooler equilibrium temperatures, since they receive a smaller integrated stellar flux, which drives hydrodynamic escape. Whilst one can explore these basic dependencies analytically (see Appendix \ref{app:analytics}), the easiest way to fully understand these effects, in conjunction with thermodynamic cooling and contraction, is with numerical models. We use the semi-analytic numerical models for XUV photoevaporation from \citet{Owen2017,OwenEstrada2020} and core-powered mass-loss from \citet{Gupta2019,Gupta2022} to numerically model populations of planets undergoing atmospheric-mass loss driven by both mechanisms \citep[see][for a full discussion of both models]{Rogers2021b}. For both models we assume an atmospheric adiabatic index of $\gamma=5/3$, a core heat capacity of $10^7$ erg g$^{-1}$ K$^{-1}$ \citep{Valencia2010} and an opacity scaling law of $\kappa \propto P^\alpha T^\beta$, where $\alpha=0.68$, $\beta=0.45$ and $\kappa = 1.29 \times 10^{-2}$ cm$^2$ g$^{-1}$ at $100$ bar and $1000$~K \citep{RogersSeager2010}. Both models rely on the hydrodynamic escape of hydrogen-dominated material, hence we expect the predicted mass-radius distributions to be very similar. 

Chronologically speaking, there are three dominant atmospheric processes that small, close-in exoplanets with H/He dominated atmospheres undergo. Firstly, atmospheric mass is accrued via core-nucleated accretion whilst immersed in a protoplanetary disc \citep[e.g.][]{Rafikov2006,Lee2014,Piso2014,Ginzburg2016}. Then, as the disc disperses, the atmospheric mass of some planets is rapidly removed through a ``boil-off'' process (also referred to as ``spontaneous mass-loss'') as the confining pressure from the disc is removed on timescales $\sim 10^5$~yrs \citep[e.g.][]{Ikomi2012,Owen2016,Ginzburg2016}. This is appropriate for smaller mass planets $\lesssim 10M_\oplus$, since larger mass cores may open gaps in the gaseous protoplanetary disc, resulting in different atmospheric evolution during disc dispersal. Finally, once the disc has completely dispersed, these latter processes transition into XUV photoevaporation and core-powered mass-loss \citep[e.g.][]{LopezFortney2013,Owen2013,Ginzburg2018,Gupta2019} combined with thermal cooling and contraction. 

Since we are not explicitly incorporating gaseous accretion and boil-off, our initial conditions must encompass such processes. To account for both scenarios, namely in which boil-off does/does not occur, we adopt two sets of initial conditions. The first scenario assumes that planets have undergone a boil-off phase during disc dispersal, for which we assume that planets host an initial atmospheric mass-fraction according to:
% \begin{equation} \label{eq:Xinit}
%     \log X_\text{init} \sim \mathcal{G} \bigg( \mu = \log X_\oplus + \gamma \log \bigg( \frac{M_\text{core}}{M_\oplus} \bigg ), \;
%     \sigma \bigg)
% \end{equation}
%where $\mathcal{G}$ represents a Gaussian distribution with mean $\mu$ and standard deviation $\sigma$ where $X_\oplus=0.01$, $\gamma=0.34$ and $\sigma=0.12$. 
\begin{equation} \label{eq:Xinit_boiloff}
    X_\text{init} = 0.01 \; \bigg( \frac{M_\text{c}}{M_\oplus} \bigg)^{0.44} 
    \; \bigg( \frac{T_\text{eq}}{1000\text{ K}}\bigg)^{0.25}, %\; \bigg( \frac{t_\text{disc}}{1 \text{ Myr}} \bigg)^{0.5},
\end{equation}
which comes from the theoretical predictions of \citet{Ginzburg2016}, which account for core accretion and boil-off. In the inference work from \citet{Rogers2023}, the authors showed that this relation can be accurately recovered from the data by inferring the correlation between core mass and atmospheric mass fraction prior to XUV photoevaporation for a sample of \textit{Kepler}, \textit{K2} and \textit{TESS} planets. %They showed that this empirically derived statistical relation is in excellent agreement with the theoretical scalings of  from . The statistical scatter $\sigma=0.12$ thus captures the variance in initial atmospheric mass for planets with different equilibrium temperatures, protoplanetary disc lifetimes and dispersal timescales. 

In the second scenario, we do not enforce a boil-off phase. There is currently uncertainty as to the details of this mechanism, as it is a non-standard escape process, particularly at high masses whereby gaps can be opened in the protoplanetary disc. In light of this, we provide an additional agnostic set of initial atmospheric mass fractions, drawn log-uniformly in the range:
\begin{equation} \label{eq:Xinit_agnostic}
    X_\text{init} \sim \mathcal{U} (10^{-3}, 0.3),
\end{equation}
where $\mathcal{U}$ is a {\jr{log-uniform distribution}}, where the lower limit avoids large mass cores hosting negligible atmospheric mass fractions (we assume planets with $X \leq 10^{-4}$ to be completely stripped i.e. super-Earths), whilst the upper limit is chosen to avoid self-gravitating atmospheres which are known to be extremely rare and the semi-analytic models do not account for \citep{WolfgangLopez2015,Rogers2023}. In essence, this distribution accounts for all possible initial atmospheric conditions. As we shall show, even this agnostic set of initial conditions accurately reproduces the observations.

We assume planetary cores are of Earth-like composition, such that they have a silicate-to-iron ratio of 67:33 \citep{Owen2017,Gupta2019,Rogers2021}, making use of the mass-radius relations from \citet{Fortney2007}. To approximately match the stellar sample from \citet{Luque2022}, we adopt a Gaussian stellar mass distribution centred at $0.3M_\odot$ with a standard deviation of $0.1M_\odot$. We evolve a population of $10^5$ planets for each mass-loss model for $5$~Gyrs to match the approximate ages of observed planets, although this final age makes no difference to the final mass-radius distribution\footnote{This is because the trend of entropy with mass is maintained across various ages, meaning that the slope and position of the mass-radius plane are age-insensitive.}. We randomly draw orbital periods from a broken power law:
\begin{equation}{\label{eq:P_distr}}
\frac{d N}{d\;\text{log}P} \propto \begin{cases}
    P^{a}, & {P < P_0\; \text{days}} \\
    P^{b}, & {P > P_0\; \text{days}},
  \end{cases}  
\end{equation}
where $a=1.0$, $b=-1.5$ and  $P_0 = 8.0$ are chosen to approximately match the population of observed planets orbiting M-dwarfs from \textit{Kepler} \citep[e.g.][]{Petigura2022}. We also place an upper limit on orbital periods of $30$ days, since most M dwarf orbiting planets with measured masses and radii are observed with \textit{TESS}, which has a baseline capable of observing planets out to this orbital separation. We randomly draw the planet core masses in a log-uniform manner, so as to evenly sample the mass-radius diagram. Finally, we remove planets with an RV semi-amplitude $\leq 30$ cm s$^{-1}$ to approximate current RV sensitivity limits.

\section{Results and Discussion} \label{sec:Results}

\subsection{Mass-radius relation for sub-Neptunes with rocky cores and H/He atmospheres} \label{sec:MR_relations}

Figure \ref{fig:MR_combined} demonstrates the mass-radius relations for a population of rocky cores, initially hosting H/He rich atmospheres, that have undergone thermal evolution and atmospheric mass-loss over 5 Gyrs. Since both photoevaporation (see left-hand panels) and core-powered mass-loss models (see right-hand panels) derive from hydrodynamic escape mechanisms, their predictions are very similar in this plane. The bimodal distribution is clearly seen, with super-Earths typically residing at orbital separations corresponding to higher equilibrium temperatures and having been stripped of their hydrogen-dominated atmospheres. As such, super-Earths fall on an Earth-like composition line in the mass-radius diagram. Sub-Neptunes, on the other hand, maintain, despite some atmospheric mass-loss, a H/He atmosphere, the amount of which scales with core mass among other variables, such that more massive cores retain larger atmospheric mass-fractions. These H/He atmospheres increase the radii of sub-Neptunes above that expected for an Earth-composition core. We note that the mass-radius relation for planets in the absence of atmospheric mass-loss i.e. constant atmospheric mass fraction \citep[e.g. see Figure 1 of][]{Lopez2014} is less steep than the H/He mass-radius relations from \citet{Zeng2019} (see Section \ref{sec:ZengComparison}). Hence, the mass-radius observations for sub-Neptunes can only be fit with H/He atmospheric mass-fractions that scale with planet mass, which is a natural outcome of the hydrodynamic atmospheric-loss processes discussed above.

In the top and middle panels, we show mass-radius distributions for both sets of initial conditions; boil-off (see Equation \ref{eq:Xinit_boiloff}) and agnostic (see Equation \ref{eq:Xinit_agnostic}) respectively. One can see that the boil-off scenario produces a narrow mass-radius distribution, whilst the agnostic initial conditions produce a wider range in sub-Neptune radii for a given mass, owing to the increased range in initial atmospheric mass fractions. {\jr{Note however that even with this set of agnostic initial conditions, the majority of sub-Neptunes sit at small radii, close to the models that started with boil-off initial conditions. This is because thermal evolution and mass-loss of H/He dominated atmospheres naturally produce this relation after Gyrs, independent of initial conditions.}}

To provide a useful reference for comparison with future observations, we quantify these mass-radius relations, which we highlight are appropriate for sub-Neptunes in the range $1.0 \lesssim M_\text{p} / M_\oplus \lesssim 30$, with quartic logarithmic functions:
\begin{equation} \label{eq:MR_NEW}
    \begin{split}
        \frac{R_\text{p}}{R_\oplus} =  a_0 & + a_1 \ln \bigg( \frac{M_\text{p}}{M_\oplus} \bigg) \; \, + a_2 \ln \bigg( \frac{M_\text{p}}{M_\oplus} \bigg)^2  \\ 
         & + a_3 \ln \bigg( \frac{M_\text{p}}{M_\oplus} \bigg)^3 + a_4 \ln \bigg( \frac{M_\text{p}}{M_\oplus} \bigg)^4,
    \end{split}
\end{equation}
where coefficients are summarised for photoevaporation and core-powered mass-loss in Tables \ref{table:1} and \ref{table:2} respectively. Since the boil-off scenario is extremely narrow (top panels of Figure \ref{fig:MR_combined}), we quantify mass-radius relations (orange dashed line) for both models by calculating median planet size for increasing bins in planet mass. We then fit these median values to Equation \ref{eq:MR_NEW}. Similarly, for the agnostic initial conditions (bottom panels of Figure \ref{fig:MR_combined}), we quantify this wider mass-radius distribution by finding $2\sigma$ limits in planet size for increasing bins in planet mass.

\begin{table}[]
\centering
\begin{tabular}{||l c c c||} 
 \hline
  & Boil-Off & Agnostic (lower) & Agnostic (upper) \\ [0.5ex] 
 \hline\hline
 $a_0$ & 1.3104 & 1.2131 & 1.5776 \\ 
 $a_1$ & 0.2862 & 0.2326 & 0.7713 \\
 $a_2$ & 0.1329 & -0.0139 & 0.5921 \\
 $a_3$ & -0.0174 & 0.0367 & -0.2325 \\
 $a_4$ & 0.0002 & -0.0065 & 0.0301 \\ [1ex] 
 \hline
\end{tabular}
\caption{Coefficients for mass-radius relations for photoevaporation, given by a quartic logarithmic equation from Equation \ref{eq:MR_NEW}. Boil-off initial atmospheric conditions (see dashed-orange line in Figure \ref{fig:MR_combined}) are from Equation \ref{eq:Xinit_boiloff}, agnostic initial atmospheric conditions (see dotted-orange lines in Figure \ref{fig:MR_combined}) are from Equation \ref{eq:Xinit_agnostic}, with upper and lower planet size bounds given.}
\label{table:1}
\end{table}

\begin{table}[]
\centering
\begin{tabular}{||l c c c||} 
 \hline
  & Boil-Off & Agnostic (lower) & Agnostic (upper) \\ [0.5ex] 
 \hline\hline
 $a_0$ & 1.3255 & 1.5776 & 1.2131 \\ 
 $a_1$ & 0.4168 & 0.7713 & 0.2326 \\
 $a_2$ & 0.1567 & 0.5921 & -0.0139 \\
 $a_3$ & -0.07224 & -0.2325 & 0.0367 \\
 $a_4$ & 0.01092 & 0.0301 & -0.0065 \\ [1ex] 
 \hline
\end{tabular}
\caption{Same as Table \ref{table:1}, but for core-powered mass-loss models.}
\label{table:2}
\end{table}

In the bottom panels of Figure \ref{fig:MR_combined}, we show our predicted mass-radius relations from Equation \ref{eq:MR_NEW} alongside the observed sample from \citet{Luque2022}, consisting of $48$ planets orbiting $26$ M-dwarfs systems with stellar masses $0.1 \lesssim M_* / M_\odot \lesssim 0.6$. Figure \ref{fig:MR_combined} clearly demonstrates that the mass-radius observations are in excellent agreement with sub-Neptunes which have rocky interiors and H/He atmospheres provided that their thermal evolution and mass-loss histories are accounted for. 

In the case of boil-off initial conditions, (orange-dashed line) the mass-radius relation from our atmospheric evolution and mass-loss models is degenerate with bodies of a 1:1 silicate-to-ice ratio \citep{Zeng2019} (blue solid line). In the case of agnostic initial conditions, (orange dotted-lines) the mass-radius relation encompasses all observed planets. Finally, we also highlight planets with blue-shaded circles that have confirmed escaping hydrogen/helium atmospheres. Namely, these are K2 18 b \citep{Benneke2019,dosSantos2020}, GJ 436 b \citep{Bean2008,Pont2009,Knutson2011,Ehrenreich2015,Turner2016}, GJ 3470 b \citep{Fukui2013,Nascimbeni2013,Crossfield2013,Dragomir2015,Awiphan2016,Bourrier2018,Ninan2020} and GJ 1214 b \citep[although we highlight that this is a tentative detection from][]{OrellMiquel2022}. K2 18 b is an interesting case, since it is close\footnote{In fact, other literature values would place K2 18 b precisely on the mass-radius relations for H/He atmospheres and water-worlds \citep[e.g.][]{Sarkis2018}.} to the mass-radius relations for atmospheric evolution (orange dashed line) and water-worlds (blue solid). However, the direct hydrogen detection suggests it is inconsistent with the water-world hypothesis since such planets cannot host significant hydrogen atmospheres whilst still being consistent with observed masses and radii. 

We also note that whilst many observed intermediate mass planets ($2\lesssim M_\text{p} / M_\oplus \lesssim 10$) are tightly clustered around the mass-radius relations for water-worlds and H/He atmospheres with boil-off initial conditions, there are many high-mass planets, including those with escaping H/He detections; GJ 3470 b, GJ 436 b, and GJ 1214 b, that sit above both of these mass-radius relations. They do however sit within the bounds of H/He mass relations with agnostic initial conditions. As discussed in Section \ref{sec:Method}, boil-off is likely inefficient for planets with $M_\text{c} \gtrsim 10 M_\oplus$ since such planets will begin to open gaps in their protoplanetary discs, hence implying the agnostic initial conditions (Equation \ref{eq:Xinit_agnostic}) are more appropriate for such planets. The observations appear to support this notion.  We highlight that more work is needed to understand boil-off and that these planets provide important tests of such processes.

% Other planets in this population are very likely to host H/He atmospheres, since this is the most plausible way to construct a planet with bulk densities consistent with their observations \citep[e.g.][]{Rogers2015,JontofHutter2016}.

\subsection{Verifying mass-radius relations with MESA} \label{sec:MESA}
Whilst the semi-analytic model of atmospheric evolution for photoevaporation from \citet{Owen2017,OwenEstrada2020} and core-powered mass-loss from \citet{Ginzburg2018,Gupta2019} are computationally inexpensive and thus allow large populations of planets to be generated, they lack complex physics such as a detailed model for convection, self-gravity and realistic equations of state. Therefore, as in \citet{Owen2017}, we corroborate our semi-analytical modelling from Figure \ref{fig:MR_combined} by comparing our results with numerical models performed with \textit{Modules for Experiments in Stellar Astrophysics} (\verb|MESA|) \citep{Paxton2011,Paxton2013,Paxton2015,Paxton2018}, which solves and evolves the stellar structure equations with accurate H/He equations of state from \citet{Saumon1995} and dust-free opacity tables from \citet{Freedman2008} for low-mass and irradiated planets. These sophisticated models remove free parameters from the problem, such as choices in adiabatic index and opacities since these are determined self-consistently. We follow previous works to model low-mass planets \citep{Owen2013,Owen2016,ChenRogers2016,Kubyshkina2020,Malsky2020}, and evolve each model for $5$~Gyrs, adopting stellar irradiation performed with the $F_*-\Sigma$ routine from \verb|MESA| \citep{Paxton2013}, which injects irradiative flux within a column density of $\Sigma$. For these models, we follow \citet{Owen2013,Owen2016} and assume $\Sigma=250 \text{ g cm}^{-2}$, appropriate for opacities to incoming stellar irradiation of $\kappa_\nu = 4 \times 10^{-3} \text{ cm}^2 \text{ g}^{-1}$ \citep{Guillot2010}.  

\begin{figure*}
	\includegraphics[width=1.7\columnwidth]{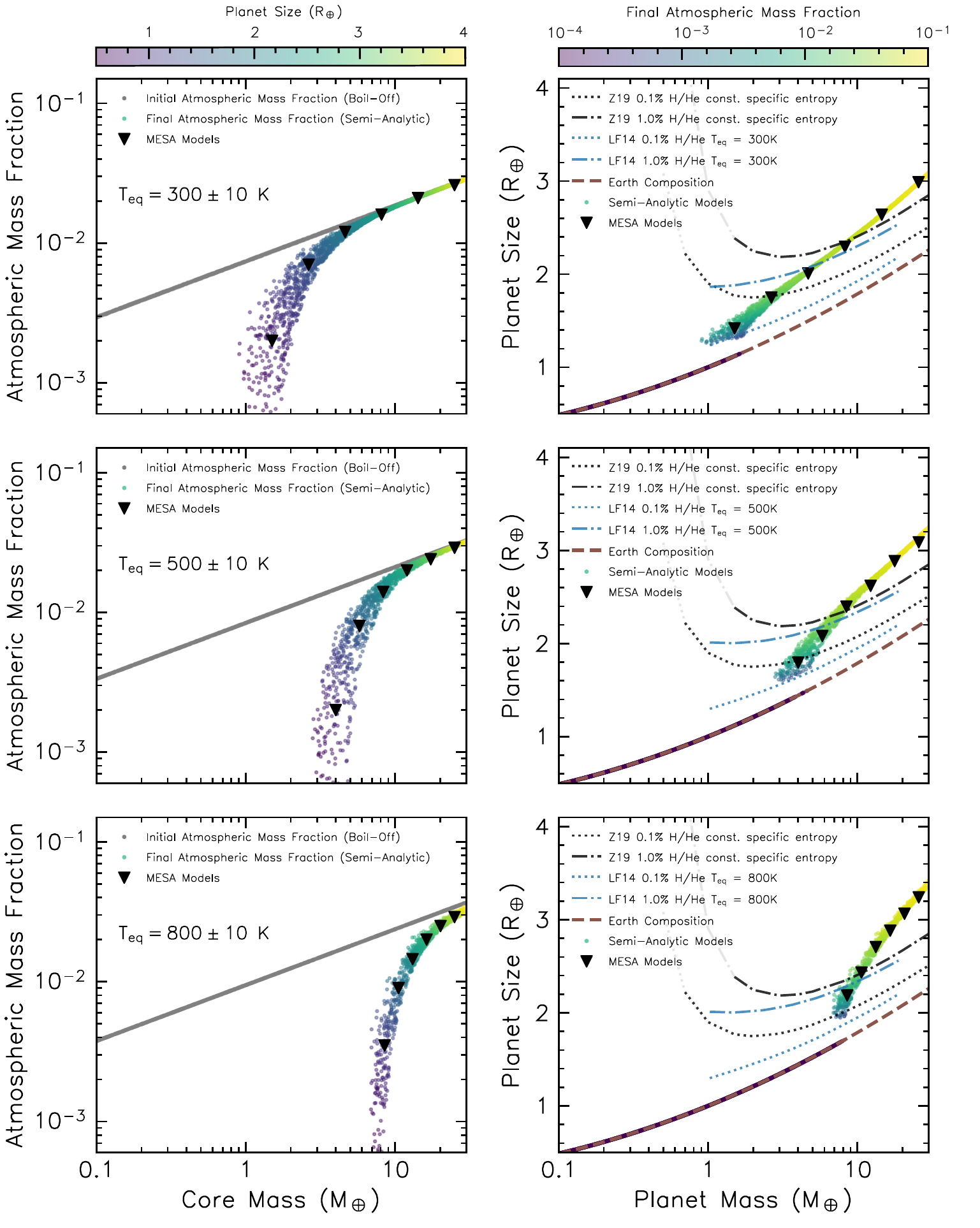}
    \centering
        \cprotect\caption{The final atmospheric mass fractions (left-hand column) and planet radii (right-hand column) after $5$~Gyr of photoevaporative evolution for populations of planets with equilibrium temperatures of $300\pm10$~K (top row), $500\pm10$~K (middle row) and $800\pm10$~K (bottom row). Colours represent planet size in the left-hand panels and final atmospheric mass fraction in the right-hand panels, demonstrating that larger atmospheric mass fractions lead to larger planets and vice versa. All planets start their evolution with an initial distribution of atmospheric mass fractions (displayed as grey points) that account for gaseous core accretion and boil-off during protoplanetary disc dispersal (see Equation \ref{eq:Xinit_boiloff}). To corroborate these semi-analytic results, we also perform numerical models with \verb|MESA|, which are shown as black triangles. In general, sub-Neptunes only exist at larger masses for higher equilibrium temperatures. The mass-radius distribution (as seen in Figure \ref{fig:MR_combined}) is a superposition of all equilibrium temperatures. In the right-hand column, we compare our mass-radius distributions with the models of \protect{\citet{Lopez2014}} in blue and \protect{\citet{Zeng2019}} in black for atmospheric mass fractions of $0.1\%$ and $1.0\%$. We highlight that the models of \protect{\citet{Zeng2019}} assume constant specific entropy defined with a temperature at fixed pressure at $100$~bar (not to be confused with the equilibrium temperature)  and therefore suggest a dramatic increase in planet radius for lower-mass planets. The models of \protect{\citet{Lopez2014}} consider irradiation and cooling for planets with constant atmospheric mass fraction, meaning they are more appropriate for analysis of planet composition. Our mass-radius models account for loss-induced scaling of atmospheric mass with planet mass, meaning they are appropriate for comparisons of planet populations in the mass-radius diagram.} \label{fig:MESA_vs_Zeng} 
\end{figure*} 

In Figure \ref{fig:MESA_vs_Zeng}, we show populations of planets evolved with \verb|MESA| at equilibrium temperatures of $300$K, $500$K and $800$K in the top, middle and bottom panels respectively, represented with black triangles. For simplicity, we only compare these results with the semi-analytic photoevaporation models since planets of different core masses can be stripped at slightly different equilibrium temperatures under the core-powered mass-loss model. For the purposes of population-level mass-radius diagrams, however, the differences between the two models are inconsequential. 

One can see from Figure \ref{fig:MESA_vs_Zeng} that the \verb|MESA| models are in excellent agreement with our adopted semi-analytic photoevaporation models which are also shown in Figure \ref{fig:MESA_vs_Zeng} for small ranges in equilibrium temperatures at $300 \pm 10$~K, $500 \pm 10$~K and $800 \pm 10$~K. Note that mass-loss is not explicitly included in these \verb|MESA| models. Instead, we adopt the final atmospheric mass-fractions from the semi-analytic photoevaporation models (as shown with black triangles in the left-hand panels of Figure \ref{fig:MESA_vs_Zeng}) and then evolve the planets in \verb|MESA| with this atmospheric mass fraction to calculate their radii after $5$~Gyrs. Since the majority of atmospheric escape under photoevaporation typically occurs in the first $\sim 100$~Myr, this is akin to beginning the \verb|MESA| simulations at the end of this period in order to accurately determine their radii at $5$~Gyrs. As Figure \ref{fig:MESA_vs_Zeng} demonstrates, these models robustly confirm the mass-radius relations found with our semi-analytic approach.

Figure \ref{fig:MESA_vs_Zeng} also highlights important points about atmospheric evolution. Firstly, in the left-hand panels, the final atmospheric mass fractions demonstrate that planets of different core masses evolve to host very different atmospheric mass fractions. For an initial boil-off distribution represented with grey points (see Equation \ref{eq:Xinit_boiloff}), low-mass planets are stripped of their atmosphere (numerically identified with an atmospheric mass fraction $\leq 10^{-4}$) whilst the highest-mass planets retain most of their initial atmospheric mass and therefore match the initial distribution. Intermediate-mass planets, however, lose progressively less atmosphere with increasing core mass. A common generalisation is that sub-Neptunes host an atmospheric mass-fraction of $\sim 1\%$ \citep{Owen2017}, since this value maximises the mass-loss timescale and naturally leads to a population of planets that retain their H/He atmosphere. Whilst this is good to an order of magnitude, Figure \ref{fig:MESA_vs_Zeng} clearly demonstrates that this an oversimplification, with larger planets naturally retaining a greater atmospheric mass fraction due to their increased gravitational potential-wells (which is also shown analytically in Appendix \ref{app:analytics}). Furthermore, this distribution changes as a function of equilibrium temperature (i.e. different rows in Figure \ref{fig:MESA_vs_Zeng}), with planets at lower equilibrium temperatures able to maintain more atmospheric mass for a given core mass. Figure \ref{fig:MESA_vs_Zeng} also demonstrates the importance of initial conditions for such planets, since high-mass planets maintain an atmospheric mass that follows their initial distribution. The population-level mass-radius diagram (as shown in Figure \ref{fig:MR_combined}) is therefore a superposition of different planets at different core masses, equilibrium temperatures and ages, with their initial conditions playing a progressively more influential role for higher masses. 

\subsection{Comparison with Zeng et. al. 2019 models} \label{sec:ZengComparison}
In the right-hand panel of Figure \ref{fig:MESA_vs_Zeng}, we compare our semi-analytic and \verb|MESA| models with numerical models of \citet{Zeng2019}, which provide mass-radius relations for rocky cores hosting a H/He atmospheric mass fraction under the assumption of constant specific entropy, defined with a temperature at a pressure of $100$ bar, although we highlight that this temperature is frequently misinterpreted as the planetary equilibrium temperature. For reference, for an adiabat with $\gamma=5/3$, set such that its temperature is $500$~K at $100$~bar, the temperature at $1$~bar is $\lesssim 80$~K, which is far below the typical planetary equilibrium temperatures currently observed. We stress that such models are not applicable to evolved sub-Neptunes to perform quantitative analysis. Examples of these mass-radius relations are shown in black-dashed lines in Figure \ref{fig:MESA_vs_Zeng} for atmospheric mass fractions of $0.1\%$ and $1.0\%$ with specific entropy set with a temperature of $500$~K at $100$~bar. These curves have a characteristic and dramatic increase in size for smaller-mass planets, which comes from the assumption of constant atmospheric mass at constant specific entropy. However, atmospheric evolution naturally allows planet atmospheres to cool and contract, with smaller-mass planets cooling more due to their reduced heat capacity. Combining this with mass-loss, which further reduces the atmospheric mass retained by smaller mass planets results in the mass-radius relations found in Figure \ref{fig:MR_combined}. We note that if one wishes to analyse an individual planet in the mass-radius diagram, then the mass-radius relations of \citet{Lopez2014} at constant age, which are also shown in Figure \ref{fig:MESA_vs_Zeng}, are more appropriate since these include the essential physical processes (cooling and irradiation for a given atmospheric mass fraction) that shape the radius of small exoplanets with hydrogen atmospheres. Alternatively, the publicly available \verb|evapmass| code from \citet{OwenEstrada2020} includes the semi-analytic atmospheric structure models adopted in this work. In the case of analysing populations of planets in the mass-radius diagram, we recommend the relations derived in this work (see Equation \ref{eq:MR_NEW}).

\subsection{Mass-radius relations for sub-Neptunes around FGK stars} \label{sec:FGK}
In this letter, we have focused on planets orbiting M dwarfs, as is the case with the observational work of \citet{Luque2022}. As we have shown, our choice of physically motivated initial conditions and ranges in equilibrium temperatures yield mass-radius relations with an intrinsic spread in planet radii for a given mass (see Figures \ref{fig:MR_combined} and \ref{fig:MESA_vs_Zeng}). There are, however, additional factors that can contribute to the mass-radius distribution spread, that we have not included in our models. As highlighted in \citet{Kubyshkina2022}, variability in high-energy stellar luminosity \citep[e.g.][]{Tu2015,Johnstone2021,Ketzer2022} can increase the range in planet sizes, since stars of different initial rotation rates will produce different X-ray/EUV flux and thus different mass-loss rates for the orbiting planets. In addition, observational uncertainties in planet radii will increase the spread in the mass-radius distribution due to purely statistical scatter.

It is interesting to note that the underlying mass-radius distribution does not significantly change when considering planets orbiting FGK stars. Although such planets will receive a larger flux at a given orbital period, we find from our mass-radius models that this bias tends to simply produce a larger ratio of super-Earth to sub-Neptune occurrence rates, since more planets can be stripped of their H/He atmosphere. Indeed, this result is consistent with the demographic work of \citet{Petigura2022}, from which one can calculate the ratio in occurrence rates of super-Earths to sub-Neptunes to find it increased, with values of $0.29\pm0.07$, $0.34\pm0.05$ and $0.54\pm0.10$ for a stellar mass bins of $[0.5,0.7] M_\odot$, $[0.7,1.1] M_\odot$ and $[1.1,1.4] M_\odot$ respectively. We do highlight, however, that there are other ways in which this ratio may increase, such as varying the core mass or orbital period distributions as a function of stellar mass.

One major difference between low and high-mass stars, however, is that transit observations (such as those from \textit{Kepler}, \textit{K2} and \textit{TESS}) can achieve a higher photometric precision around M dwarfs due to their smaller stellar radii and hence larger $R_\text{p} / R_*$. Such surveys are also biased to observe planets within a smaller range in equilibrium temperatures since the transit probability of planets at large orbital periods (and therefore low equilibrium temperatures) decreases rapidly. Different mission targeting strategies also change the observed population e.g. \textit{Kepler} was sensitive to planets with orbital periods $\sim 100$ days, but specifically targeted FGK stars, whereas \textit{TESS} currently targets nearby bright stars (and is therefore biased to M-dwarfs), with sensitivity out to orbital periods $\sim 30$ days. These arguments taken together suggest that the \textit{observed} mass-radius distribution around M-dwarfs is expected to have less scatter compared to that for planets around FGK stars. This is indeed the case when comparing Figures 1 and S19 from \citet{Luque2022} \citep[see also Figure 12 from][for an example of a synthetic mass-radius distribution for FGK stars in the presence of bias and measurement uncertainty]{Rogers2021}. We find that the underlying mass-radius distribution, in the absence of statistical scatter and bias\footnote{We also highlight that the bias of planet mass measurements is currently not quantifiable, since such surveys are not based on homogeneous observations of a well-defined sample of stars.} (as summarised by Equation \ref{eq:MR_NEW} under different initial conditions), is approximately the same across FGKM spectral types but that the relative occurrence of super-Earths with respect to sub-Neptunes increases for more massive (luminous) stellar types.

\section{Discussion and conclusions} \label{sec:Conclusions}

In this letter, we calculate mass-radius relations of small, close-in exoplanets that host H/He dominated atmospheres with self-consistent, physically motivated evolution models, which are summarised for photoevaporation and core-powered mass-loss models in Eq. \ref{eq:MR_NEW} and Tables \ref{table:1} and \ref{table:2} respectively. We consider two sets of initial conditions; the first, in which planets undergo a boil-off phase, whereby a large fraction of atmospheric mass is lost during protoplanetary disc dispersal \citep[e.g.][]{Ikomi2012,Owen2016,Ginzburg2016}, which yields a relatively tight mass-radius relation after mass-loss and thermal evolution. In the second scenario, we adopt agnostic initial conditions (see Equation \ref{eq:Xinit_agnostic}), which yields a larger spread in final radii. These relations (see Figure \ref{fig:MR_combined}) incorporate thermodynamic cooling, atmospheric escape and stellar irradiation and are therefore suited for compositional analyses of populations of sub-Neptunes in the mass-radius diagram \citep[e.g][]{Wu_Lithwick2013,Weiss2014,HaddenLithwick2014,Rogers2015,Dressing2015,Wolfgang2016,Chen2017,VanEylen2021}. We show that accounting for atmospheric mass-loss yields left-over atmospheric mass fractions that scale with planet mass i.e. larger planets retain a larger fraction of their total mass in hydrogen and show that these results give an excellent match to the mass and radius measurements of sub-Neptunes in \citet{Luque2022}, independently of assumed initial conditions. In addition, we show that the boil-off initial conditions yield a mass-radius relation that is completely degenerate with that corresponding to a 1:1 silicate-to-ice ratio. We note that our study moves beyond the H/He mass-radius relations of \citet{Zeng2019}, as demonstrated in Figure \ref{fig:MESA_vs_Zeng}, which assume a constant specific entropy for constant atmospheric mass factions as a function of planet mass. Such models are therefore not applicable to planets undergoing atmospheric evolution. 

In \citet{Luque2022}, a sample of observed exoplanets orbiting M-dwarfs is used to argue that planets with rocky interiors and H/He atmospheres cannot explain the observed cluster of planets around the 1:1 silicate-to-ice ratio compositional line in the mass-radius diagram. In Figure \ref{fig:MR_combined} we have presented our new mass-radius relations for small planets with hydrogen atmospheres (see Equation \ref{eq:MR_NEW}) and show that they are in fact completely consistent with the data, once thermal evolution and mass-loss are properly accounted for. A strong degeneracy therefore still exists between the water-world and silicate/iron-hydrogen models. We find that planets with different equilibrium temperatures and atmospheric masses for a given core mass yield a natural spread in the mass-radius relation (see Figure \ref{fig:MR_combined}) that does not vary dramatically for different stellar types. We do note that other factors that we have not taken into account, such as high-energy stellar luminosity variability \citep{Kubyshkina2022} and observational uncertainty will act to increase the spread in the sub-Neptune mass-radius relation. Nevertheless, we note that many high-mass planets $\gtrsim 10 M_\oplus$ in the sample from \citet{Luque2022}, including GJ 436 b, GJ 3470 b and GJ 1214, which have confirmed escaping H/He atmospheric detections, sit well above the mass-radius relations for both water-worlds and hydrogen atmosphere models that assume an initial boil-off scenario. We speculate that such planets were less susceptible to boil-off \citep{Owen2016,Ginzburg2016} due to their increased mass (potentially due to gap-opening in their protoplanetary discs) and therefore entered the XUV photoevaporation/core-powered mass-loss phase with larger atmospheric mass fractions. We highlight that further work is needed to understand this important stage in exoplanet evolution.

In light of the results shown in Figure \ref{fig:MR_combined}, we corroborate the well-known result that a planet's mass and radius alone are often insufficient to break its internal composition degeneracy \citep{Valencia2007,Rogers2010}. Probing for hydrogen and helium presence around low-mass planets with spectroscopic observations is one promising avenue \citep[e.g.][]{Ehrenreich2015,Lavie2017,Bourrier2018,Yan2018,Spake2018,dosSantos2020,Ninan2020,Zhang2022} although we highlight that a non-detection in hydrogen Ly-$\alpha$ does not necessarily indicate the lack of a hydrogen-dominated atmosphere \citep{Owen2023}. Moreover, observations from \textit{JWST} may provide insights into the abundance of H$_2$O in high mean-molecular weight atmospheres of sub-Neptunes and thus the prevalence of water-worlds. In this letter, we have also analysed the occurrence rates from \citet{Petigura2022} to find that the ratio of super-Earths to sub-Neptunes increases with increasing stellar mass. Since larger mass stars produce larger luminosities, this result tentatively supports the notion that stellar irradiation is key in evolving sub-Neptunes into super-Earths via atmospheric escape. 

In addition, if one can accurately measure planet age, then one can determine how planets and the observed radius gap, separating the super-Earths from the sub-Neptunes, evolves with time. Under atmospheric evolution models, the radius gap is expected to evolve on $\sim 100$ Myr to Gyr timescales \citep[][]{Gupta2020,Rogers2021} since hydrogen-dominated atmospheres dramatically change in size as they cool due to their low mean-molecular weight. Water-worlds on the other hand will not significantly change in size after formation since their sizes are dominated by their ice-silicate composition and not H/He dominated atmospheres. Indeed this demographic analysis has been performed in the works of \citet[e.g.][]{Berger2020b,Sandoval2021,Chen2022} to show that the radius gap evolves on $\sim 100$ Myr to Gyr timescales. Moreover, a recent study from \citet{Fernandes2022} calculated occurrence rates for a sample of exoplanets around young stars. They find tentative evidence for the decrease in sub-Neptune size with stellar age, which is indicative of significant cooling and contraction of low mean-molecular-weight H/He dominated sub-Neptunes with time, although we note that presently this sample size is small. A larger sample of planets with accurate ages may shed light on this issue. Furthermore, mass measurements of young sub-Neptunes may show that such planets are indeed inflated and therefore extremely under-dense H/He-rich proto-sub-Neptunes, destined to cool and contract to the evolved population we observe today \citep{Owen2020}.

{\jr{Finally, we highlight that, in this work, we have adopted a definition of water worlds to match that of \citet{Zeng2019,Luque2022} i.e. 1:1 silicate-to-ice ratio. In reality, one can relax this condition to consider planets hosting H/He atmospheres \textit{and} water \citep[i.e.][]{Venturini2016,Lambrechts2019,Mordasini2020,Venturini2020}, which may be in the form of sequestered H$_2$O \citep[e.g.][]{Dorn2021,Vazan2022} or steam atmospheres \citep[e.g.][]{Aguichine2021,Kimura2020}. On this note, we highlight the work of \citet{Lopez2017} that demonstrates that the escape of steam atmospheres is not efficient enough to reproduce the exoplanet demographics. We also stress that the observed lack of planets in the radius gap \citep[e.g.][]{VanEylen2018,Ho2023} places strong constraints on the composition spread allowed in such models i.e. if one allows too much spread in water-mass fraction, then one cannot reproduce the emptiness of the radius gap. We conclude that whilst water is likely present in many, if not all sub-Neptunes to some degree, clear evidence for a population of water worlds with large ice-mass fractions (such as 1:1 silicate-to-ice ratios) still remains elusive.}}

%% Also note that the akcnowlodgment environment does not support long amounts of text. If you have a lot of people and institutions to acknowledge, do not use this command. Instead, create a new \section{Acknowledgments}.
\section*{Acknowledgements}
{\jr{We would like to kindly thank the anonymous reviewer for comments that improved this letter.}} We would also like to thank Akash Gupta, Ruth Murray-Clay, Erik Petigura and Vincent Van Eylen for discussions that helped improve this work. JGR is supported by the Alfred P. Sloan Foundation under grant G202114194 as part of the AEThER collaboration. HES gratefully acknowledges support from NASA under grant number 80NSSC21K0392 issued through the Exoplanet Research Program. JEO is supported by a Royal Society University Research Fellowship. This project has received funding from the European Research Council (ERC) under the European Union’s Horizon 2020 research and innovation programme (Grant agreement No. 853022, PEVAP). For the purpose of open access, the authors have applied a Creative Commons Attribution (CC-BY) licence to any Author Accepted Manuscript version arising.

\section*{Software}
{\jr{The inlists and scripts used in \verb|MESA| calculations are available on Zenodo under an open-source Creative Commons Attribution license: \dataset[doi:10.5281/zenodo.7647801]{https://doi.org/10.5281/zenodo.7647801}.}}

\bibliography{references}{}
\bibliographystyle{aasjournal}

\appendix
\section{Analytic Arguments} \label{app:analytics}
A natural outcome of atmospheric mass-loss is that larger planetary cores at a cooler equilibrium temperature retain a larger atmospheric mass fraction due to their increased gravitational potential well and reduced irradiation. It is this basic result that introduces a slope in the radius-period valley rather than having a single radius that separates H/He-rich planets from stripped cores. Further, as we demonstrate in Figures \ref{fig:MR_combined} and \ref{fig:MESA_vs_Zeng}, if one takes this simple fact into account, the mass-radius relation of planets that have undergone atmospheric escape naturally reproduces the observations. Here we demonstrate this analytically, as well as highlighting why this result holds for both photoevaporation and core-powered mass-loss, emphasising that both models agree with the exoplanet demographics because the underlying physics is similar. To analytically show that larger planetary cores at cooler equilibrium temperatures retain a larger atmospheric mass fraction, consider a planet with core mass $M_\text{c}$ and radius $R_\text{c}$, equilibrium temperature $T_\text{eq}$, hosting a H/He dominated atmosphere. This atmosphere is split into a convective interior (assumed to be adiabatic with index $\gamma$) and a radiative exterior (assumed to be isothermal, with a temperature equal to $T_\text{eq}$). The location of the radiative-convective boundary is $R_\text{rcb}$ with density $\rho_\text{rcb}$. Following from \citet{Owen2017}, the mass of the convective interior scales as:
\begin{equation} \label{eq:Matm}
    M_\text{atm} \propto R_\text{rcb}^3 \, \rho_\text{rcb} \bigg( \frac{R_\text{B}}{R_\text{rcb}} \bigg)^{\frac{1}{\gamma-1}} \, I_2 \bigg( \frac{R_\text{c}}{R_\text{rcb}}, \gamma \bigg ),
\end{equation}
where $R_\text{B} = GM_\text{c} / 2 c_\text{s}^2$ is the Bondi radius for isothermal sound speed $c_\text{s} = (k_\text{B} T_\text{eq} / \mu m_\text{H})^{1/2}$. Here, $I_2$ is a dimensionless integral which accounts for the mass distribution within the atmosphere:
\begin{equation} \label{eq:I2}
    I_2 \bigg( \frac{R_\text{c}}{R_\text{rcb}}, \gamma \bigg ) = \int^1_{R_\text{c}/R_\text{p}} x^2 \bigg(\frac{1}{x} - 1 \bigg)^{\frac{1}{\gamma-1}} dx.
\end{equation}
In the case of hydrodynamic escape of planetary atmospheres, the mass-loss rate $\dot{M}$ scales as:
\begin{equation} \label{eq:MassLossRate}
    \dot{M} \propto R_\text{rcb}^2 \, \rho_\text{rcb} \, c_\text{s} \, \mathcal{M}_\text{rcb},
\end{equation}
where $\mathcal{M}_\text{rcb}$ is the Mach number of the escaping flow, evaluated at the radiative-convective boundary. For an isothermal outflow, the Mach number is only a function of $R_\text{B} / R_\text{rcb}$ and given by:
\begin{equation}
    \mathcal{M}_\text{rcb} = \sqrt{-W_0 \bigg [ - \bigg( \frac{R_\text{B}}{R_\text{rcb}} \bigg)^4 \exp{ \bigg(-C-4\frac{R_\text{B}}{R_\text{rcb}}} \bigg ) \bigg ]},
\end{equation}
where $W_0$ is the real branch of the Lambert W function \citep[see][]{Cranmer2004} and $C$ is a constant. It is through this constant $C$ that photoevaporation and core-powered mass-loss distinguish themselves. This is because even in photoevaporation, the outflow is approximately isothermal between the radiative-convective boundary and XUV penetration point. Thus, the Mach number at the radiative-convective boundary is still given by a solution to the isothermal flow problem. However, unlike core-powered mass-loss where the outflow is approximately isothermal all the way to the sonic point (and hence is the classic Parker wind solution - \citealt{Parker1958}), in the photoevaporative case the outflow must typically supply a higher mass-loss rate to the XUV penetration point than provided by the Parker wind solution, and is necessarily faster. Thus, in the case of XUV photoevaporation, $C<-3$ to provide these more powerful outflows, whereas for core-powered mass-loss, $C=-3$ \citep[see][]{Lamers1999}. For planets that have maintained a significant mass in H/He i.e. sub-Neptunes, one can state that their mass-loss timescale $t_\text{loss} = M_\text{atm} / \dot{M}$, will be approximately constant. Hence, combining Equations \ref{eq:Matm} and \ref{eq:MassLossRate}, one finds that:
\begin{equation}
    t_\text{loss} = \frac{M_\text{atm}}{\dot{M}} \propto \frac{R_\text{rcb} \,\bigg( \frac{R_\text{B}}{R_\text{rcb}} \bigg)^{\frac{1}{\gamma-1}} \, I_2 }{c_\text{s} \, \mathcal{M}_\text{rcb}} \propto \text{const.}
\end{equation}
This expression is dominated by the exponential term within $\mathcal{M}_\text{rcb}$, and only varies logarithmically with $C$, $R_\text{rcb}$ and $I_2$ \footnote{This can be seen by Taylor expansion of the Lambert W function in the limit $R_{\rm rcb}/R_B\ll1$.}. Thus, since the variation is only logarithmic with $C$, there is no leading order difference in the dependence on fundamental parameters between photoevaporation and core-powered mass-loss\footnote{Note that while the scaling on fundamental parameters is the same, the normalisation constant varies between the two models as photoevaporation drives more powerful outflows, but for a shorter period of time.}. Hence, one can state that:
\begin{equation} \label{eq:RbRrcbConst}
    \frac{R_\text{B}}{R_\text{rcb}} \propto \frac{M_\text{c}}{R_\text{rcb} c_\text{s}^2} \propto \text{const.}
\end{equation}
Now, by combining Equations 8, 9 and 11 from \citet{Owen2017}, \citep[see also][]{Ginzburg2016}, which assume radiative diffusion at the radiative-convective boundary for a cooling/Kelvin-Helmholtz timescale $\tau_\text{KH}$, one finds that the density at the radiative-convective boundary scales as:
\begin{equation} \label{eq:RhoRCB}
    \rho_\text{rcb} \propto \frac{R_\text{rcb} T_\text{eq}^3 \tau_\text{KH}}{M_\text{atm} \kappa} \bigg( \frac{I_2}{I_1} \bigg),
\end{equation}
where $\kappa$ is the opacity and $I_1$ is another dimensionless integral accounting for the binding energy of the planet:
\begin{equation} \label{eq:I1}
    I_1 \bigg( \frac{R_\text{c}}{R_\text{rcb}}, \gamma \bigg ) = \int^1_{R_\text{c}/R_\text{p}} x \bigg(\frac{1}{x} - 1 \bigg)^{\frac{1}{\gamma-1}} dx.
\end{equation}
Combining the density at the radiative-convective boundary from Equation \ref{eq:RhoRCB} with the atmospheric mass from Equation \ref{eq:Matm}, and noting that $R_\text{B} / R_\text{rcb}$ is approximately constant from Equation \ref{eq:RbRrcbConst}, one can show that:
\begin{equation}
    \frac{M_\text{c}}{R_\text{rcb} c_\text{s}^2} \propto X^\frac{1}{2} \, M_\text{c}^{-\frac{1}{2}} \, T_\text{eq}^{\frac{1}{4}} \, \kappa^{\frac{1}{4}} I_1^\frac{1}{4} \, I_2^{-\frac{1}{2}}.
\end{equation}
where the atmospheric mass fraction is defined as $X \equiv M_\text{atm} / M_\text{c}$ and we have assumed that for a set of planets with the same age, their Kelvin-Helmholtz timescale will be approximately constant. Finally, recalling again that $M_\text{c} / R_\text{rcb} c_\text{s}^2$ is approximately constant from Equation \ref{eq:RbRrcbConst}, one finds that:
\begin{equation}
    X \propto M_\text{c} \, T_\text{eq}^{-\frac{1}{2}} \, \kappa^{-\frac{1}{2}} I_1^{-\frac{1}{2}} \, I_2.
\end{equation}
If one numerically evaluates the dimensionless integrals $I_1$ and $I_2$ \citep[see Figure 11 from][]{Owen2017}, one can show that $I_1^{-0.5} I_2$ is approximately constant as a function of $R_\text{c} / R_\text{rcb}$. If one also assumes that the opacity $\kappa$ is constant, then one finally finds that the atmospheric mass fraction of planets that have undergone mass-loss scales approximately linearly with core mass and inversely with the square root of equilibrium temperature. Moreover, this analytic argument is agnostic with respect to mass-loss models i.e. photoevaporation vs. core-powered mass-loss. The main takeaway result is that larger-mass planets at cooler equilibrium temperatures will retain larger atmospheric mass fractions if they have undergone mass-loss. This result is key to explaining the mass-radius distribution of exoplanets. Note however from Figure \ref{fig:MESA_vs_Zeng}, that whilst final atmospheric mass fraction increases with core mass at a given equilibrium temperature, it is not a linear or even log-linear relation. This is the case when one fully evaluates the integrals of Equations \ref{eq:I2} and \ref{eq:I1} and takes non-constant opacities into account, as is the case in the semi-analytic models of \citet{Owen2017, OwenEstrada2020} and \citet{Ginzburg2018,Gupta2019}.



\end{document}